\documentclass[sigconf]{aamas} 

\usepackage{balance} 

\setcopyright{ifaamas}
\acmConference[AAMAS '21]{Proc.\@ of the 20th International Conference on Autonomous Agents and Multiagent Systems (AAMAS 2021)}{May 3--7, 2021}{Online}{U.~Endriss, A.~Now\'{e}, F.~Dignum, A.~Lomuscio (eds.)}
\copyrightyear{2021}
\acmYear{2021}
\acmDOI{}
\acmPrice{}
\acmISBN{}

\title{A Framework for Integrating Gesture Generation Models into Interactive Conversational Agents}
\subtitle{Demonstration Track}

\author{Rajmund Nagy}
\authornote{Both authors contributed equally to the paper.}
\affiliation{
  \institution{KTH, Stockholm, Sweden}
  }

\author{Taras Kucherenko}
\authornotemark[1]
\affiliation{
  \institution{KTH, Stockholm, Sweden}
  }
 
\author{Birger Moell}
\affiliation{
  \institution{KTH, Stockholm, Sweden}
  }

\author{André Pereira}
\affiliation{
  \institution{KTH, Stockholm, Sweden}
  }

\author{Hedvig Kjellstr{\"o}m}
\affiliation{
  \institution{KTH, Stockholm, Sweden}
  }

\author{Ulysses Bernardet}
\affiliation{
  \institution{Aston University, Birmingham, UK}
  }

\begin{abstract}
Embodied conversational agents (ECAs) benefit from non-verbal behavior for natural and efficient interaction with users. Gesticulation -- hand and arm movements accompanying speech -- is an essential part of non-verbal behavior. Gesture generation models have been developed for several decades: starting with rule-based and ending with mainly data-driven methods.  To date, recent end-to-end gesture generation methods have not been evaluated in a real-time interaction with users. We present a proof-of-concept framework, which is intended to facilitate evaluation of modern gesture generation models in interaction. 

We demonstrate an extensible open-source framework that contains three components: 1) a 3D interactive agent; 2) a chatbot backend; 3) a gesticulating system. 
Each component can be replaced, making the proposed framework applicable for investigating the effect of different gesturing models in real-time interactions with different communication modalities,  chatbot backends, or different agent appearances. The code and video are available at the project page \href{https://nagyrajmund.github.io/project/gesturebot}{https://nagyrajmund.github.io/project/gesturebot}.
\end{abstract}


\newcommand{\BibTeX}{\rm B\kern-.05em{\sc i\kern-.025em b}\kern-.08em\TeX}

\begin{document}


\pagestyle{fancy}
\fancyhead{}


\begin{teaserfigure}
  \centering
  \includegraphics[width=0.6\linewidth]{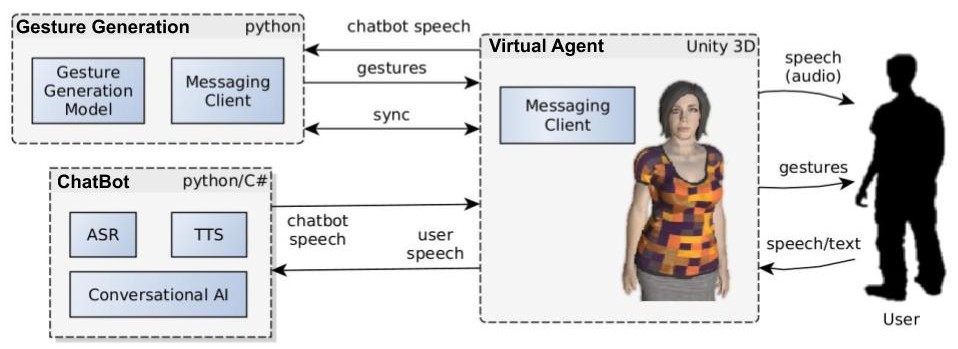}
  \caption{Architecture of the framework for integrating gesture generation models.}
  \label{fig:architecture}
  \Description[A schematic depiction of the framework's software architecture and the communication between the components.]{A schematic depiction of the framework's software architecture and a user. The three components are: 1) Gesture Generation (in python), containing a Gesture Generation Model and a Messaging Client; 2) ChatBot (in python or C#), containing an ASR module, a TTS module and a conversational AI; 3) a Virtual Agent (in Unity 3D), with the Messaging Client integrated into it. The communication between the components is shown with arrows as follows. The user interacts with the virtual agent through speech or text, the Virtual Agent sends the user speech to the ChatBot, which sends the ChatBot speech back. The Virtual Agent sends this ChatBot speech to the Gesture Generation component, which returns the synthesized gestures. Finally, the Virtual Agent acts out the gestures towards the user.}
\end{teaserfigure}

\maketitle 



\section{Introduction}
Humans use non-verbal behavior to signal their intent, emotions and attitudes \citep{knapp2013nonverbal,Nixon2018}.
Similarly, embodied conversational agents (ECAs) can be more engaging when having appropriate nonverbal behavior~\cite{salem2011friendly}. It is therefore desirable to enable conversational agents to communicate nonverbally.

Currently existing implementations of ECAs rely primarily on pre-recorded animations or require handcrafted specification of the motion \cite{kopp2004synthesizing, lhommet2015cerebella, rodriguez2016singing}, e.g. in XML-based formats \cite{kopp2006towards}.  However, recent developments in the field of gesture generation make it possible to produce realistic gestures in a purely data-driven fashion \cite{alexanderson2020style, kucherenko2021moving, yoon2019robots}. To date, many of these recent gesture generation methods have not been evaluated in a real-time interaction with users. A potential reason for the lack of evaluation in interaction for the recent models is the difficulty of setting up an interactive conversational agent.

In this work, we outline a framework for embedding data-driven gesture generation models into conversational agents. We envision that this framework with accelerate development of interactive embodied agents with end-to-end gesticulation models.




Our framework is modular, which enables it to be used for a wide range of scientific investigations about intelligent virtual agents, such as experimenting with their voices, gestures, breathing, conversational complexity or gender. For our demonstration,  we integrate the speech- and text-driven model developed by Kucherenko et al.~\cite{kucherenko2020gesticulator} into an ECA built with Unity, and we show the flexibility of our approach by demonstrating our framework with two different chatbot backends.
\section{System description}

Our open source system is composed of a 3D virtual agent in Unity, a chatbot backend with text-to-speech capabilities and a neural network that generates gesturing motion from speech (Figure~\ref{fig:architecture}). The communication between the components consists of sending audio, text or motion file in a message; in our implementation, this is facilitated by the open-source Apache ActiveMQ message broker\footnote{\url{https://activemq.apache.org/}} and the STOMP protocol\footnote{\url{https://stomp.github.io/}}.

Each component is replaceable and is described in the corresponding section below. 

\subsection{Virtual agent in Unity}

We provide the virtual environment and the user interface as a Unity scene. The end user interacts with the conversational agent through voice input or a text field.

By using Unity, we ensure that the system can be easily extended with new modules by other researchers in the future. Furthermore, it makes it possible to tailor the environment and the character model according to the requirements of the application (e.g., explore virtual/mixed reality applications).

\subsection{Chatbot backend}

%

The user's input message is sent to the chatbot backend that produces the agent's response as text and audio.
The chatbot backend is comprised of a speech recognition system, a neural conversational model and a text-to-speech synthesizer. For our demonstration, we present two implementations of this component.

In the first configuration, Google's popular DialogFlow platform \cite{sabharwal2020introduction} is used with its automatic speech recognition module to enable voice-based interaction with the agent. The interfacing to DialogFlow is implemented in a C\# module inside Unity.

In the second configuration, we adapt the open-domain chatbot BlenderBot \cite{roller2020recipes} and a text-to-speech model called Glow-TTS \cite{kim2020glow} to build a virtual agent with free-form conversation capabilities. We leverage open-source implementations (provided by HuggingFace \cite{wolf-etal-2020-transformers} and Mozilla TTS\footnote{\url{https://github.com/mozilla/TTS}}) to seamlessly integrate the two models into the Python backend.

\subsection{Gesture generation model}
Based on the output of the chatbot backend, the gesture generation model synthesizes the corresponding motion sequence. We adapt a recent gesture generation model called Gesticulator \cite{kucherenko2020gesticulator},
which is an autoregressive neural network that takes acoustic features combined with semantic information as its input, and generates the corresponding gesticulation as a sequence of upper-body joint angles, which is a widely used representation in computer animation and robotics. 

In the original paper \cite{kucherenko2020gesticulator}, the network was trained on the Trinity Speech-Gesture dataset \cite{ferstl2018investigating}, consisting of 244 minutes of speech and motion capture recordings of spontaneous monologues acted out by a male actor. The input features -- log-power spectrograms for audio and BERT \cite{devlin2018bert} word embeddings for text -- are extracted and concatenated at the frame level, and a 1.5 s (30 frames at 20 FPS) sliding window of input features is used for predicting every motion frame, motivated by gesture-speech alignment research.

We tailor the base Gesticulator model to our interactive agent with the following adjustments:

\begin{enumerate}
    \item We replace the audio features from spectrograms to the extended Geneva Minimalistic Acoustic Parameter Set \cite{eyben2015gemaps}, normalized to zero mean and unit variance. We qualitatively found that it results in better motion for synthesized voice. 
    \item The text transcriptions that are used for training the model contain precise word timing information, which is usually not available in real-time settings. Therefore, when interacting with a user, we approximate it with speech utterance lengths that are proportional to the syllable count.
    \item Finally, we replace the BERT word embedding with FastText~\cite{bojanowski2017enriching} (which has significantly lower dimensionality) in order to reduce the feature extraction time.
\end{enumerate}

\section{Limitations}
At the current stage of development, each of the components  has some important limitations:
\begin{enumerate}
    \item Both available chatbot backends introduce several
    seconds of processing time before the agent responds to the user, which might currently affect immersion in the interaction.
    \item The synthesized voice of the agent yields out-of-distribution audio samples which significantly degrade the quality of the generated motion. This could be improved by replacing audios in the dataset with synthetic audios \cite{sadoughi2016head} or by training a TTS model on the audio from a speech-gesture dataset  \cite{alexanderson2020generating}.
    \item Finger motion is not modelled by the gesture generation model due to poor data quality in the dataset.
\end{enumerate}

However, the system's modular design of replaceable components allows addressing these limitations in the future. For instance, it is straightforward to replace Gesticulator with a model that generates full-body motion or to change the chatbot backend as shown in our two distinct examples with DialogFlow and Blenderbot.

\section{Conclusions and future work}
We have presented a framework for integrating state-of-the-art data-driven gesture generation models with embodied conversational agents. As highlighted by the GENEA gesture generation challenge \cite{kucherenko2021large}, the gesture generation field needs a reliable benchmark. The proposed framework provides such a possibility; it can be used to compare gesture generation models in real-time interactions. 

There are many directions in which this work can be extended in the future. More diverse gesticulation can be achieved by choosing a probabilistic gesture generation model instead of a deterministic one like Gesticulator. Incorporating stylistic control \cite{alexanderson2020style, ferstl2020understanding} in ECA to allow expression of different emotions
is a promising direction. Moreover, the proposed framework can be used in user studies to
investigate the human perception of different gesticulation.






\newpage


\balance
\bibliographystyle{ACM-Reference-Format} 
\bibliography{refs.bib}



\end{document}